\documentclass[usenatbib,onecolumn]{mn2e}
\usepackage{graphicx}
\usepackage{times}
\newcommand{\beq}{\begin{equation}}
\newcommand{\eeq}{\end{equation}}
\newcommand{\bea}{\begin{eqnarray}}
\newcommand{\eea}{\end{eqnarray}}
\def\lsim{\raise0.3ex\hbox{$\;<$\kern-0.75em\raise-1.1ex\hbox{$\sim\;$}}}
\def\gsim{\raise0.3ex\hbox{$\;>$\kern-0.75em\raise-1.1ex\hbox{$\sim\;$}}}
\title
{Diagnosing GRB Prompt Emission Site with Spectral Cut-Off Energy} 
\author[Nayantara Gupta and Bing Zhang]
{Nayantara Gupta\thanks{nayan@physics.unlv.edu} and 
Bing Zhang\thanks{bzhang@physics.unlv.edu}\\ 
Department of Physics and Astronomy, 
University of Nevada Las Vegas, Las Vegas, NV 89154, USA}
\begin{document}
\date{Accepted 2007; Received 2007; in original form 2007}
\pagerange{\pageref{firstpage}--\pageref{lastpage}} \pubyear{2007}
\maketitle
\label{firstpage}
\begin{abstract}
The site and mechanism of gamma-ray burst (GRB) prompt emission is still
unknown. Although internal shocks have been widely discussed as the
emission site of GRBs, evidence supporting other emission sites, including 
the closer-in photosphere where the fireball becomes transparent and 
further-out radii near the fireball deceleration radius where magnetic 
dissipation may be important, have been also suggested recently.
With the successful operation of the GLAST experiment, prompt high
energy emission spectra from many GRBs would be detected in the
near future. We suggest
that the cut-off energy of the prompt emission spectrum from a GRB 
depends on both the fireball bulk Lorentz factor and the unknown
emission radius from the central engine. If the bulk Lorentz factor
could be independently measured (e.g. from early afterglow observations),
the observed spectral cutoff energy can be used to diagnose the emission
site of gamma-rays. This would provide valuable information to understand 
the physical origin of GRB promp emission.
\end{abstract}

\section[]{Introduction}
The emission of photons in the prompt phase of a GRB may last from
less than a second to hundreds of seconds (see e.g. \citet{mesz4} for
a recent general review on GRBs). The exact physical mechanism and the
emission site of the observed prompt
GRB emissions are still unknown \citep{zhang1,zhang0}. The
physical processes which may lead to this emission include synchrotron/jitter 
emission, inverse Compton scattering or a combination of thermal and
non-thermal emission components. Internal shocks have been widely 
discussed in the literature as the possible emission site of GRB prompt emission
(Rees \& M\'esz\'aros 1994; M\'esz\'aros et al. 1994; 
Kobayashi et al. 1997; Daigne \& Mochkovitch
1998; Pilla \& Loeb 1998; Panaitescu \& M\'esz\'aros 2000; Lloyd \&
Petrosian 2000; Zhang \&
M\'esz\'aros 2002a; Dai \& Lu 2002; Pe'er \& Waxman 2004, 2005; Pe'er
et al. 2005; Gupta \& Zhang 2007).
Within this model the radius of emission from the central engine ($r$) is 
related to the bulk Lorentz factor
($\Gamma$) and the variability time ($t_v$) through
$r=\Gamma^2ct_v$. For typically observed values of bulk Lorentz factor
$\Gamma\sim 300$ and variability time $t_v\sim 0.01~$sec the emission
radius is $r \sim 3\times 10^{13}~$cm. However, the internal shock
origin of GRB prompt emission is not conclusive. The baryonic or pair 
photospheres of the GRB fireball have been argued to be another possible
emission site of prompt GRB emission (e.g. Thompson 1994; M\'esz\'aros
\& Rees 2000; M\'esz\'aros et al. 2002; Ryde 2005). This model has the
merit of potentially reproducing some observed empirical correlations among 
GRB prompt emission properties (e.g. Rees \& M\'esz\'aros 2005; Ryde et al. 
2006; Thompson et al. 2007; cf. Zhang \& M\'esz\'aros 2002). 
On the other hand, an analysis of Swift
early afterglow data led Kumar et al. (2007) to conclude that the
prompt emission site is between $10^{15}$-$10^{16}$cm from the central
engine. This emission site is too large for typical internal shocks
but too small for external shocks\footnote{This radius can be still
accommodated within the internal shock picture if the typical variability 
time scale is a significant fraction of the burst duration. 
Liang et al (2006) discovered that if the steep decay segment observed in
Swift X-ray afterglows is due to the curvature effect of the high latitude
emission with respect to the line of sight (Kumar \& Panaitescu 2000; 
Zhang et al. 2006), the required time zero point $t_0$ usually leads the 
beginning of the steep decay ($t_p$), and $(t_p-t_0)$ (effectively the 
variability time scale for the internal shock scenario) is a significant 
fraction of the burst duration. The internal shock scenario therefore can be 
still consistent with Kumar et al.'s analysis.}. Emission at this radius may 
be related to magnetic dissipation (e.g. Spruit et al. 2001; Drenkhahn
\& Spruit 2002; Zhang \& M\'esz\'aros 2002; Giannios \& Spruit 2007).
In both of the above two non-internal shock models for GRB prompt 
emissions, it is possible to argue that the emission radius $r$ could be
in principle related to the Lorentz factor $\Gamma$ and variability $t_v$ in
a non-trivial (e.g. other than $r=\Gamma^2 ct_v$) manner.
For example, in the photosphere
model, $r$ is defined by the optically-thin condition, and is not directly
related to $t_v$ which is related to the time history of the GRB 
central engine. In the magnetic dissipation model, if the energy 
dissipation occurs locally (i.e. the emission region scale is much 
smaller than the emission radius), it is possible to have $r>\Gamma^2
ct_v$. In general, it is reasonable to treat $r$ as an independent
quantity with respect to $\Gamma$ and $t_v$.

High energy photons produced in the prompt emission region 
are expected to interact with
lower energy photons before escaping as a result of
two photon attenuation. In general, the internal optical depth of
$\gamma\gamma$ interactions depends on $\Gamma$, $t_v$, and the radius 
of the emission region. Traditionally, internal shocks have been 
taken as the default model of GRB prompt emission, and the pair 
attenuation optical depth has been expressed as a function of $\Gamma$
and $t_v$ only (e.g. Piran 1999; Lithwick \& Sari 2001). 
The pair attenuation process is expected to leave a cutoff spectral
feature in the prompt emission spectrum, and detecting such a spectral
cutoff by high energy missions such as GLAST has been discussed as
an important method to estimate the bulk Lorentz factor $\Gamma$
of the fireball \citep{math1,math2}. 
The issue of unknown emission radius $r$ makes the picture more
complicated.  It is no longer straightforward to
estimate $\Gamma$ with an observed spectral cutoff energy. On the
other hand, there are other independent methods of estimating
$\Gamma$ using early afterglow (e.g. Sari \& Piran 1999; 
Zhang, Kobayashi \& M\'esz\'aros 2003) or prompt emission 
(Pe'er et al. 2007) data and there have been cases of such measurements 
(Molinari et al. 2007; Pe'er et al. 2007). It is then
possible to use the observed spectral cutoff energy to diagnose the 
unknown GRB emission site if $\Gamma$ is measured by other means. 
In this paper we release the internal shock assumption
and re-express the cutoff energy more generally as a function of
$r$ and $\Gamma$. We then discuss an approach of diagnosing the
GRB prompt emission site using the future cutoff energy data retrieved
by GLAST and other missions. Lately Murase \& Ioka (2007, see also
Murase \& Nagataki 2006) also 
independently discussed to use the pair cutoff signature to diagnose
whether the emission site is the pair/baryonic photosphere. We discuss 
this topic more generally to diagnose any emission site.
We emphasise that our method can be used to constrain $r$ only when a 
clear cut-off is observed in the high energy photon spectrum from a GRB.

\section[]{Parametrization of Internal Optical Depth}
The cross section of two-photon interaction can be generally expressed as 
(Gould \& Schreder 1967)
\bea
\sigma_{\gamma_h\gamma_l}(E_{\gamma_h}^{\prime},E_{\gamma_l}^{\prime},\theta')=
\frac{3}{16}\sigma_T(1-{b}^2)\Big[(3-{b}^4)\ln{\frac
{1+{b}}{1-{b}}}-2{b}(2-{b}^2)\Big]~,
\label{cross_sec}
\eea
where $\sigma_T$ is the Thomson cross section, $b=
[1-(E_{\gamma_l,th}^{\prime}/E_{\gamma_l}^{\prime})]^{1/2}$ is the center 
of mass dimensionless speed of the pair produced, $E_{\gamma h}^{'}$,
$E_{\gamma l}^{'}$ and $\theta'$ are the high- and low-energy photon
energies and their incident angles in the comoving frame of the GRB
ejecta.  The threshold energy of pair 
production for a high energy photon with energy $E_{\gamma_h}^{\prime}$ is 
\beq
E_{\gamma_l,th}^{\prime}=\frac{2(m_ec^2)^2}{E_{\gamma_h}^{\prime}(1-\cos{\theta'})}~.
\eeq
In the comoving frame, the relative velocity of the high energy 
and low energy photons along
the direction of the former is $c(1-\cos{\theta'})$. For an isotropic
distribution the fraction of low energy photons moving in the
differential cone at an angle between $\theta'$ and $(\theta'+d\theta')$ is
$\frac{1}{2}\sin\theta'd{\theta'}$.  The inverse of the mean free path for
$\gamma_h$ $\gamma_l$ interactions
$l_{\gamma_h\gamma_l}^{-1}(E_{\gamma_h}^{\prime})$ can be calculated as
\beq
l_{\gamma_h\gamma_l}^{-1}(E_{\gamma_h}^{\prime})=\frac{1}{2}\int_{-1}^{+1} 
d(\cos{\theta'})(1-\cos{\theta'})l_{\gamma_h\gamma_l\theta'}^{-1}
(E_{\gamma_h}^{\prime},\theta')
\label{path_l}
\eeq
where
\bea
l_{\gamma_h\gamma_l\theta'}^{-1}(E_{\gamma_h}^{\prime},\theta')=
\int_{E_{\gamma_l,th}^{\prime}}^{\infty} d{E_{\gamma_l}^{\prime}}
\frac{dn_{\gamma_l}(E_{\gamma_l}^{\prime})}{d{E_{\gamma_l}^{\prime}}}
\sigma_{\gamma_h\gamma_l}(E_{\gamma_h}^{\prime},E_{\gamma_l}^{\prime},\theta')~,
\eea
$\frac{dn_{\gamma_l}(E_{\gamma_l}^{\prime})}{dE_{\gamma_l}^{\prime}}$
is the specific number density of low energy photons of the GRB in the comoving
frame. The observed low energy photon spectrum (per unit energy per unit
area) $\frac{dN_{\gamma_l}^{\rm o}(E_{\gamma_l}^{\rm o})}{dE_{\gamma_{l}}^{\rm o} dA}$
from a GRB pulse can be used to estimate
$\frac{dn_{\gamma_l}(E_{\gamma_l}^{\prime})}{dE_{\gamma_l}^{\prime}}$, i.e.
\beq
\frac{dn_{\gamma_l}(E'_{\gamma_{l}})}{dE'_{\gamma_{l}}}=\frac{d_{z}^2}{r^2 
\Delta^{\prime}}\frac{dN_{\gamma_l}^{\rm o}(E_{\gamma_{l}}^{\rm o})}{dE_{\gamma_{l}}^{\rm o} 
dA}\frac{dE_{\gamma_{l}}^{\rm o}}{dE'_{\gamma_{l}}}
\label{phot_den}
\eeq 
where $\Delta^{\prime}$ is the comoving width of the shell at the radius $r$
from the central engine. We notice that the expression of $\Delta^{\prime}$ 
is function of radius (e.g. Zhang \& M\'esz\'aros 2002b): $\Delta'=r$ for
$r<r_c$; $\Delta'=r_c=\Gamma c t_v$ for $r_c \leq r<r_{s}$; and $\Delta'=r/\Gamma$
for $r \geq r_{s}=\Gamma^2 c t_v$. 
%We therefore do not specify the expression of $\Delta^{\prime}$.
Throughout the paper the superscript ``o'' denotes
the quantities measured in the observer's rest frame. 

The comoving distance of the source is 
\beq
d_z=\frac{c}{H_0}\int_{0}^{z}\frac{dx}{\sqrt{\Omega_{\Lambda}+\Omega_{m}
(1+x)^3}}~,
\eeq
which is related to the luminosity distance through $d_L=d_z(1+z)$, where $z$ 
is the redshift of the source.
Here $H_0=71 {\rm km~ s^{-1}~ Mpc^{-1}}$ 
$\Omega_{\Lambda}=0.73$ and $\Omega_{m}=0.27$ are adopted in our
calculations.  The observed
fluence is usually a broken power law with a break energy of the order
of MeV (Band et al. 1993). We assume that the two spectral indices below 
and above the break energy are $-\beta_1$ and $-\beta_2$, respectively,
usually with $\beta_1 \sim 1$ and $\beta_2 \geq 2$. 
We model the observed photon flux as
\beq
\frac{dN_{\gamma_l}^{\rm o}(E_{\gamma_{l}}^{\rm o})}{dE_{\gamma_{l}}^{\rm o} dA}=f^{\rm o}
\left\{\begin{array}{l@{\quad \quad}l} {E_{\gamma_l}^{\rm o}}^{-\beta_1} 
& E_{\gamma_l}^{\rm o}<E_{\gamma,br}^{\rm o}\\
{E_{\gamma,br}^{\rm o}}^{\beta_2-\beta_1}{E_{\gamma_{l}}^{\rm o}}^{-\beta_2} 
& E_{\gamma_l}^{\rm o}>E_{\gamma,br}^{\rm o}
\label{obs_f}
\end{array}\right.
\eeq
The break energy in the photon spectrum in the observer's frame is denoted by
$E_{\gamma,br}^{\rm o}$.
We define a new variable following the procedure discussed in Gould
and Schreder (1967) to reduce the number of integrals 
\beq
s=\frac{E_{\gamma_l}^{\prime}E_{\gamma_h}^{\prime}(1-\cos{\theta'})}{2(m_ec^2)^2}=
\frac{E_{\gamma_l}^{\prime}}{E_{\gamma_l,th}^{\prime}}=s_0 \Theta
\eeq
with
$s_0=\frac{E_{\gamma_l}^{\prime}E_{\gamma_h}^{\prime}}{(m_ec^2)^2}$, and $\Theta=
\frac{1}{2}(1-\cos{\theta'})$. As $b=(1-1/s)^{1/2}$, the pair production 
cross section can be expressed as a function of $s$. It is then possible 
to write Eq.(\ref{path_l}) as 
\bea
l_{\gamma_h\gamma_l}^{-1}(E_{\gamma_h}^{\prime})=\frac{3}{8}\sigma_T\Big(\frac{m_e^2c^4}
{E_{\gamma_h}^{\prime}}\Big)^2\int_{\frac{m_e^2c^4}{E_{\gamma_h}^{\prime}}}^{\infty} 
{E_{\gamma_l}^{\prime}}^{-2}\frac{dn_{\gamma_l}(E_{\gamma_l}^{\prime})}{dE_{\gamma_l}
^{\prime}}  Q[{s_0(E_{\gamma_l}^{\prime})}] dE'_{\gamma l}
\label{path_l2}
\eea
where
\beq
Q[s_0(E_{\gamma_l}^{\prime})]=\int_{1}^{s_0(E_{\gamma_l}^{\prime})} s\sigma(s)ds~,
\label{del_s0}
\eeq
and
$\sigma(s)=\frac{16}{3}\frac{\sigma_{\gamma_h\gamma_l}}{\sigma_{T}}$.
For moderate values of $s$ we use $\sigma(s)\simeq 1$ and the
expressions for $Q[s_0(E_{\gamma_l}^{\prime})]$ becomes $(s_0^2-1)/2$.
Substituting for $Q[s_0(E_{\gamma_l}^{\prime})]$ in
Eq.(\ref{path_l2}) we derive the final expression for
$l_{\gamma_h\gamma_l}^{-1}(E_{\gamma_h}^{\prime})$. Finally, the internal
optical depth $\tau_{int}(E_{\gamma_h}^{\rm o})$ is the ratio of width of
$\gamma_h\gamma_l$ interaction region and the mean free path of their
interaction. In most cases, this is simply
\beq
\tau_{int}(E_{\gamma_h}^{\rm o})=\triangle^{\prime} / l_{\gamma_h \gamma_l}
(E_{\gamma_h}^{\rm o})~.
\label{tau}
\eeq
In this case, comparing Eqs.(\ref{phot_den}) and (\ref{tau}) suggests that
the concrete expression of $\Delta^{\prime}$ does not enter the 
problem since it is cancelled out in the expression of 
$\tau_{int}(E_{\gamma_h}^{\rm o})$. In the photosphere models that invoke a 
continous wind from the central engine \citep{mesz2,gian1}, however, this
expression should be modified as 
\beq
\tau_{int}(E_{\gamma_h}^{\rm o})=r/ [\Gamma l_{\gamma_h \gamma_l}
(E_{\gamma_h}^{\rm o})]~.
\label{tau2}
\eeq
We therefore also consider such a case.

In the following we discuss three cases. The analytical expressions below are
only valid for Eq.(\ref{tau}). For the case of continuous photosphere models
(Eq.[\ref{tau2}]), the analytical expressions for $r<r_{is}$ are more 
complicated, and we only present the numerical results in Fig.1, where 
$\Delta'=\Gamma c t_v$ is used. \\
{\bf Case (I)}: Both the cutoff energy $E_{\gamma c}^{\rm o}$ (defined by 
$\tau_{int}(E_{\gamma_c}^{\rm o})=1$ at which energy the observed spectrum
significantly deviates from the power-law extension of the low energy
spectrum) and the threshold energy $E_{\gamma_l,th}^{\rm o}$ (for $E_{\gamma c}^{\rm o}$)
are above the break energy in the observed photon spectrum $E_{\gamma,br}^{\rm o}$, 
i.e. $E_{\gamma,br}^{\rm o}<E_{\gamma_l,th}^{\rm o}<E_{\gamma_c}^{\rm o}$ (or 
$E_{\gamma,br}'<E_{\gamma_l,th}'<E_{\gamma_c}'$). The expression
for the optical depth is the simplest for this case
\beq
\tau_{int}(E_{\gamma_h}^{\rm o})=\frac{A_1(E_{\gamma_h}^{\rm o})}{r^2}
\Big(\frac{\Gamma}{1+z}\Big)^{2-2\beta_2}
\label{opt1}
\eeq 
where 
\beq
A_1(E_{\gamma_h}^{\rm o})=\frac{3\sigma_T{d_z}^2f^{\rm o}_1}{8({\beta_2}^2-1)}
\Big(\frac{E_{\gamma_h}^{\rm o}}{{m_{e}}^2c^4}\Big)^{\beta_2-1}
\label{A1}
\eeq
and $\frac{dN_{\gamma_l}^{\rm o}(E_{\gamma_{l}}^{\rm o})}{dE_{\gamma_{l}}^{\rm o}
dA}=f^{\rm o}_1{E_{\gamma_{l}}^{\rm o}}^{-\beta_2}$ with
$f^{\rm o}_1=f^{\rm o}{E_{\gamma,br}^{\rm o}}^{\beta_2-\beta_1}$ has been assumed
for $E_{\gamma_l}^{\rm o}>E_{\gamma,br}^{\rm o}$.  In the case of
internal shocks the radius of prompt emission is $r=\Gamma^2 c
t_{v}^{\rm o}/(1+z)$ where $t_{v}^{\rm o}$ is the observed variability time scale. 
Substituting this expression of $r$ in
Eq.(\ref{opt1}) one gets
$\tau_{int}(E_{\gamma_h}^{\rm o})\propto \Gamma^{-2-2\beta_2}$,
which is consistent with \citet{lithwick}.
Our expression is more generic with $r$ being a free parameter.\\
{\bf Case (II)}: If the cutoff energy is still above the break energy, but
the threshold energy for pair production is below the break energy, i.e.
($E_{\gamma_l,th}^{\rm o}<E_{\gamma,br}^{\rm o}<E_{\gamma_c}^{\rm o}$), the expression for
internal optical depth is more complicated. Making use 
of Eq.(\ref{obs_f}), one gets
\beq
\tau_{int}(E_{\gamma_h}^{\rm o})=\frac{A_2(E_{\gamma_h}^{\rm o})}{r^2}
\Big(\frac{\Gamma}{1+z}\Big)^{2-2\beta_1}
\label{opt2}
\eeq
with
\bea
A_2(E_{\gamma_h}^{\rm o})=\frac{3\sigma_T {d_z}^2f^{\rm o}}{16}
\Big(\frac{{m_e}^2c^4}{E_{\gamma_h}^{\rm o}}\Big)^2\Big[\Big(\frac{E_{\gamma_h}^{\rm o}}
{{m_e}^2c^4}\Big)^2 \Big(\frac{E_{\gamma_h,br}^{\rm o}}{{m_e}^2c^4}\Big)^{\beta_1-1}
\Big(\frac{\beta_1-\beta_2}{(\beta_2-1)(\beta_1-1)}\Big)
\nonumber\\+{\Big(\frac{E_{\gamma_h,br}^{\rm o}}{{m_e}^2c^4}\Big)^{1+\beta_1}}
\frac{\beta_2-\beta_1}{(1+\beta_2)(1+\beta_1)}
+\Big(\frac{E_{\gamma_h}^{\rm o}}{{m_e}^2c^4}\Big)^{1+\beta_1}\frac{2}{{\beta_1}^2-1}\Big]
\label{A2}
\eea
for $\beta_1 \neq 1$, 
where $E_{\gamma_h,br}^{\rm o}$ is the energy of the high energy photons that interact
with the break-energy photons at the threshold condition, which is
defined by $E_{\gamma_h,br}^{\rm o} E_{\gamma,br}^{\rm o}=
\Big(\frac{\Gamma}{1+z}\Big)^2{m_e}^2c^4$. Equation (\ref{A2}) can be reduced
to Eq.(\ref{A1}) when $\beta_1=\beta_2$. 
For $\beta_1=1$, $\tau_{int}$ does not depend on $\Gamma$, and one has 
\beq
\tau_{int}(E_{\gamma_h}^{\rm o})=\frac{3\sigma_Tf^{\rm o}}{16}\Big(\frac{d_z}{r}\Big)^2
\Big(\frac{{m_e}^2c^4}{E_{\gamma_h}^{\rm o}}\Big)^2\Big[\Big(\frac{E_{\gamma_h}^{\rm o}}
{{m_e}^2c^4}\Big)^2\Big[\ln\Big(\frac{E_{\gamma_h}^{\rm o}}{E_{\gamma_h,br}^{\rm o}}\Big)
+\frac{1}{\beta_2-1}\Big]+\Big(\frac{E_{\gamma_h,br}^{\rm o}}{{m_e}^2c^4}\Big)^2
\Big[\frac{1}{2}-\frac{1}{1+\beta_2}\Big]-\frac{1}{2}\Big(\frac{E_{\gamma_h}^{\rm o}}
{{m_e}^2c^4}\Big)^2\Big]~.
\label{opt4}
\eeq
In this case $\Gamma$ is not needed to infer $r$.\\   
{\bf Case (III)}: In more extreme cases, usually with a low enough Lorentz factor,
one could have the cutoff energy below the break energy, i.e.
$E_{\gamma_l,th}^{\rm o}<E_{\gamma_c}^{\rm o}<E_{\gamma,br}^{\rm o}$.

In this regime, we still use Eq.(\ref{path_l2}) to calculate the internal optical
depth, but effectively one can place the upper
limit of the integration as the break energy, since 
above the break energy the photon flux falls off rapidly. This gives
\beq
\tau_{int}(E_{\gamma_h}^{\rm o})=\frac{A_3(E_{\gamma_h}^{\rm o})}{r^2}
\Big(\frac{\Gamma}{1+z}\Big)^{2-2\beta_1}
\label{opt3}
\eeq
where
\beq
A_3(E_{\gamma_h}^{\rm o})=\frac{3\sigma_T{d_z}^2f^{\rm o}}{16}\Big[\frac{2}{{\beta_1}^2-1}
\Big(\frac{E_{\gamma_h}^{\rm o}}{{m_{e}}^2c^4}\Big)^{\beta_1-1}-\frac{1}{\beta_1-1}
\Big(\frac{E_{\gamma_h,br}^{\rm o}}{m_e^2c^4}\Big)^{\beta_1-1}+\frac{1}{1+\beta_1}
\Big(\frac{m_e^2c^4}{E_{\gamma_h}^{\rm o}}\Big)^2\Big(\frac{E_{\gamma_h,br}^{\rm o}}{m_e^2c^4}
\Big)^{1+\beta_1}]
\label{A3}
\eeq

To summarize all three cases, the radius of the prompt emission can be calculated 
in terms of cutoff energies by making the internal optical depths 
(Eqs.[\ref{opt1},\ref{opt2},\ref{opt3}]) to unity
\beq
r=[A_i(E_{\gamma_c}^{\rm o})]^{1/2}\Big(\frac{\Gamma}{1+z}\Big)^{1-\beta_j}
\label{rad}
\eeq
where $j=2$ for $i=1$, and $j=1$ for $i=2,3$. 
In practice, from the observed low energy photon spectrum it is possible
to derive the low energy photon spectral parameters (including $\beta_1$,
$\beta_2$, $E_{\gamma,br}^{\rm o}$, $E_{\gamma_c}^{\rm o}$, etc.) and to identify
one applicable case among the three cases discussed.  
If the burst redshift $z$ is measured from afterglow observations, and if
the GRB bulk Lorentz factor is measured or constrained independently with
other methods (e.g. Zhang et al. 2003; Molinari et al. 2007; Pe'er et al.
2007), one can estimate the GRB emission radius using Eq.(\ref{rad}).
For the case (I) and if $\beta_2\sim 2$, a very simple expression of $r$ is 
available according to Eq.(\ref{opt1})
\beq
r=\frac{d_L}{m_ec^2\Gamma}\Big(\frac{\sigma_Tf^{\rm o}_1E_{\gamma_c}^{\rm o}}{8}\Big)^{1/2}~ 
\label{rad_1}~ 
\eeq
which could be used to quickly estimate $r$ with the data.

\begin{figure*}
\begin{center}
\includegraphics[width=7cm,height=5cm]{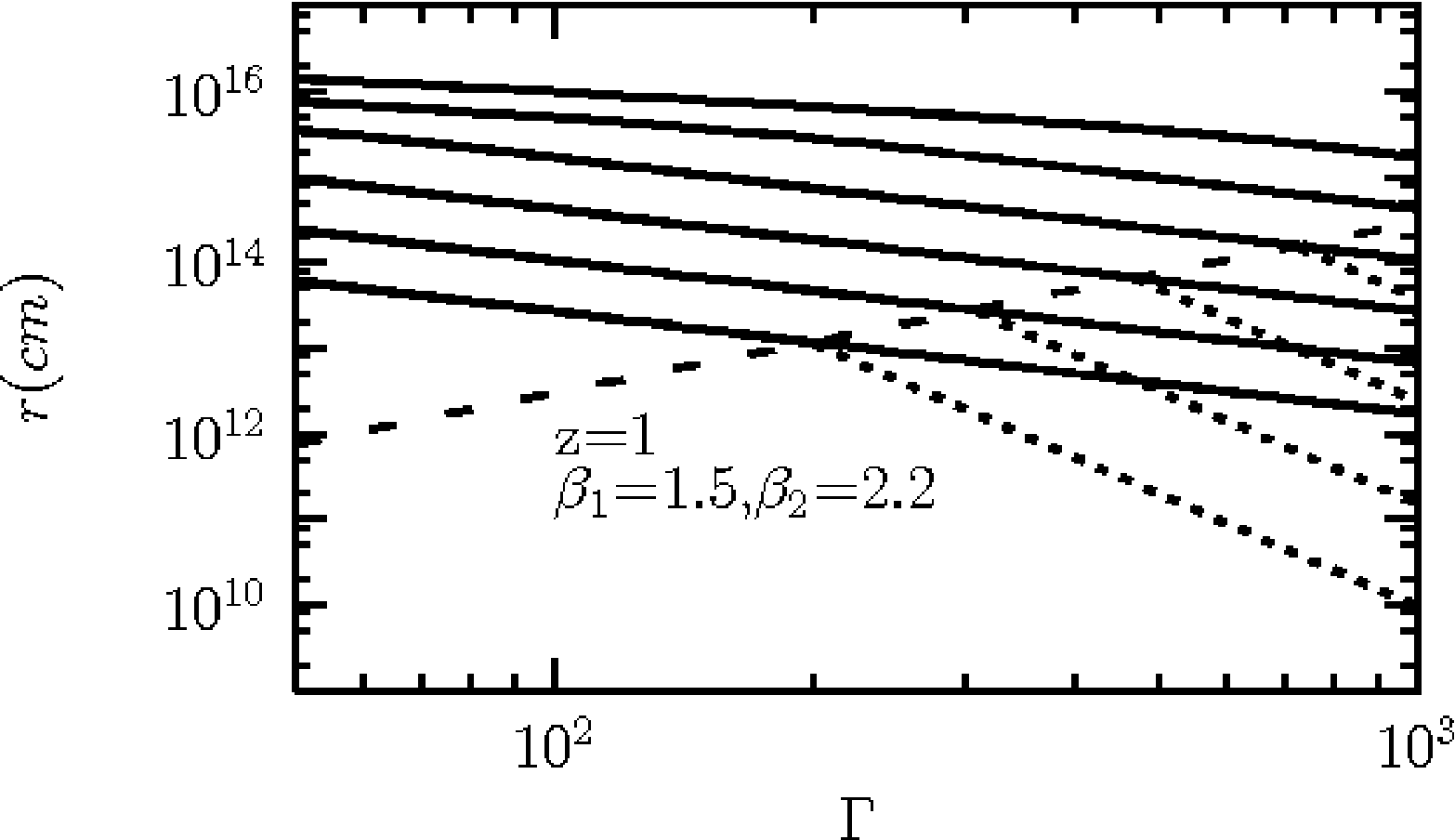}
\includegraphics[width=7cm,height=5cm]{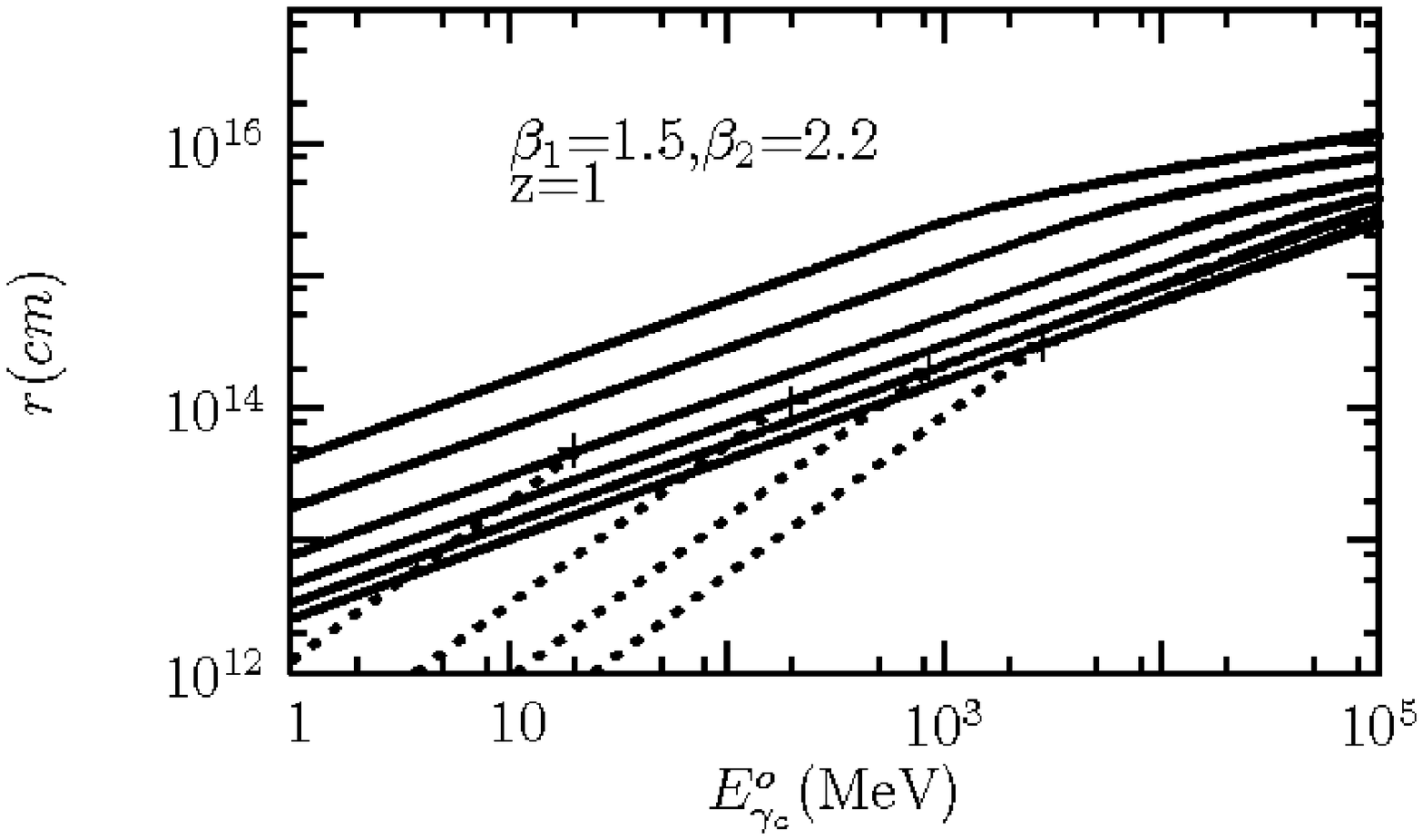}
\caption{(a) Contours of $E_{\gamma_c}^{\rm o}$ in the $r-\Gamma$ plane. From top to bottom,
$E_{\gamma_c}^{\rm o}=$50GeV, 5GeV, 0.5GeV, 50MeV, 5MeV and 0.5MeV, respectively. Model
inputs: $E_{\gamma,br}^{\rm o}=1$MeV, observed flux is $0.6{\rm MeV^{-1}cm^{-2}}$ 
at $E_{\gamma,br}^{\rm o}$, $z=1$, $\beta_1=1.5$ and $\beta_2=2.2$. The solid lines
are derived from Eq.(\ref{tau}). The dotted lines are derived from Eq.(\ref{tau2})
for $r<r_{is}$. The dashed line represents the internal shock 
$r-\Gamma$ relation, and the variability time $t_v=0.01$sec is assumed.
(b) Inferring $r$ from the observed cutoff energy. From top to bottom, $\Gamma=
100, 200, 400, 600, 800$, and 1000, respectively. Other parameters and line styles
are the same as (a). The crosses on the lines for $\Gamma=400$ to $1000$ are for 
the internal shock model with $t_v=0.01$sec.}
\end{center}
\end{figure*}

\section{Case Study and Discussion}
We have generalized the expression of the cutoff energy of prompt GRB spectrum
to include the dependences of both $\Gamma$ and $r$. We suggest that the information
of this cutoff energy is useful to diagnose the unknown location of gamma-rays.
We discuss three cases, and derive a general expression of $r$ (Eq.[\ref{rad}]). 
In Fig.1(a) we present the contours of $E_{\gamma_c}^{\rm o}$ in the $r-\Gamma$ plane
(detailed model parameters are listed in the figure caption). It is straightforward
to see that $E_{\gamma_c}^{\rm o}$ carries the information of both $\Gamma$ and $r$,
and unless the internal shock model is assumed, one cannot constrain $\Gamma$
by $E_{\gamma_c}^{\rm o}$. Figure 1(b) indicates that by knowing $\Gamma$ from other
measurements, one could diagnose $r$ with the $E_{\gamma_c}^{\rm o}$ information.
GLAST will be launched in early 2008, and LAT on board will be able to measure
the pair cutoff signature for many GRBs. With the coordinated observations with
Swift and ground-based follow up observations (to obtain $z$ and $\Gamma$ 
information), one may be able to more precisely diagnose the emission site of
gamma-rays, which is so-far subject to debate.

We notice that the maximum observed photon energy ($e_{\rm max}$)
mentioned in the paper by \citet{lithwick} is not the same as the
cut-off energy ($E_{\gamma_c}^o$) discussed in our paper.  They discussed the
maximum observed energy defined by the detector bandpass and sensitivity,
and used that energy to derive the {\em lower limit} of $\Gamma$. Within
that context, they discussed two possible limits defined by pair attenuation
(limit A) and Compton scattering by pairs (limit B), respectively. Here we
discuss the physical cutoff energy, so we can infer the actual value (not
lower limits) of $\Gamma$ and $r$ (our new addition). By definition, by 
regarding $E_{\gamma_c}^o$ as $e_{\rm max}$ in \citet{lithwick}, their 
$e_{\rm max,an}$ is just $e_{\rm thick}$, and their self-annihilation 
energy $e_{\rm self,an}$ is always smaller than or at most equal to 
$e_{\rm max}$ by definition (otherwise photons cannot be attenuated at
$e_{\rm max}=E_{\gamma_c}^o$). As a result, the limit B discussed by
\citet{lithwick} is never relevant in our context.

We have assumed that the photon field is isotropic in the comoving frame.
In some models (e.g. Lyutikov \& Blandford 2003; Thompson et al. 2007)
the emitters are moving fast in the comoving frame of the bulk flow.
This will introduce anisotropy of the photon field in the comoving frame.
Suppose the relative bulk Lorentz factor of the emitter in the comoving
frame is $\Gamma'_e \sim$ several, the photon interaction angle is at 
most $2/\Gamma'_e$. This will reduce the optical depth for pair production.
The inferred $r$ (given a same $\Gamma$) should be smaller. For example,
for $\Gamma'_e \sim 2$ the optical depth is lower by a factor of $\sim$16 and
the inferred $r$ decreases by a factor of $\sim$4.

A possible pair attenuation exponential cutoff signature may have been
observed in the pulse 2 of GRB 060105 with the joint Swift-Konus-Wind data
(Godet et al. 2007). 
The observed cutoff is around 600 keV, and the spectral index before the
cutoff is flat: $\beta_1=0.67$. The observed photon number flux at 1 MeV
is 0.5 ${\rm photons~cm^{-2}~s^{-1}~{MeV}^{-1}}$, and the observed duration 
of the pulse is about 20sec. A pseudo redshift $z\sim 4$ is inferred,
which is consistent with the broad pulse profile in the observed lightcurves.
This burst likely belongs to Case (III) discussed above. If the cutoff
feature is real and is indeed due to pair attenuation, the requirement 
that $E_{\gamma_l,th}^{\rm o}$ should be smaller than $E_{\gamma,c}^{\rm o}$ 
demands $\Gamma \leq 6$. Using Eq.(\ref{opt3}), (\ref{A3}), and  (\ref{rad}), 
with $\beta_1=0.67$, one can 
estimate $r \geq 10^{16}$ cm, which is consistent with the conclusion
drawn from analyzing the Swift X-ray data (Kumar et al. 2007). Since the
quality of the data is poor, we look forward to the high quality data 
retrieved by GLAST to finally pin down $r$ in the future.\\

We thank the anonymous referee for important remarks and Olivier Godet, Kohta Murase, 
Enrico Ramirez-Ruiz for helpful discussion/comments. This work is supported by 
NASA under grants NNG06GH62G, NNX07AJ66G and NNX07AJ64G.

\end{document}